\documentclass[proceedings, preprint]{rmaa}

\usepackage{paralist}

\usepackage{psfrag,color}

\SetYear{2009}
\SetConfTitle{THE INTERFEROMETRIC VIEW OF HOT STARS}

\title{INTERACTING BINARY STARS ENVIRONMENTS \\ AND 
THE W~SER -- DPV -- ALGOL CONNECTION} 

\author{
R. E. Mennickent,\altaffilmark{1} 
  and Z. Ko{\l}aczkowski,\altaffilmark{1,2}}

\altaffiltext{1}{Departamento de Astronom\'\i{}a, Univ. de Concepci\'on, Casilla 160-C,
Concepci\'on,  Chile
(rmennick@astro-udec.cl). 
}
\altaffiltext{2}{Instytut Astronomiczny Uniwersytetu Wroc{\l}awskiego, Kopernika 11, 51--622 Wroc{\l}aw, Poland   (zibi@astro-udec.cl).}

\shortauthor{Mennickent \& Ko{\l}aczkowski}
\shorttitle{Interacting binary star environments}

\listofauthors{R. E. Mennickent \& Z. Ko{\l}aczkowski}
\indexauthor{Mennickent, R. E.}
\indexauthor{Ko{\l}aczkowski, Z.}

\abstract{
Recent work on some kinds of interacting binaries is summarized,  
with emphasis on Cataclysmic Variables, Algol-like variables and the recently discovered Double Periodic Variables (DPVs). 
The sequence W~Serpentids
(very massive with irregular  variability and large mass loss) $\rightarrow$ DPVs (less massive with regular variability and ciclic mass loss) $\rightarrow$ Algols (even less massive with small mass loss) could correspond to an evolutionary sequence, and illustrate  the importance of the mass transfer rate in shaping  observable and mass loss properties for these systems.}

\resumen{Revisamos el trabajo reciente en algunos campos de binarias interactuantes con \'enfasis en Variables Catacl\'\i{}smicas, tipo Algol y Variables de Doble Periodo (DPVs). La secuencia de sistemas W Serpentis 
(muy masivas con variabilidad irregular y gran p\'erdida de masa) $\rightarrow$ DPVs (menos masivas con variabilidad regular  y p\'erdida de masa c\'\i{}clica) $\rightarrow$ Algoles (a\'un menos masivas con poca p\'erdida de masa) podr\'\i{}a corresponder a una secuencia evolutiva e ilustrar la importancia de la tasa de intercambio de masa en determinar las propiedades observables y de p\'erdida de masa de estos sistemas.
}

\addkeyword{Binaries}
\addkeyword{Stars: Evolution}
\addkeyword{Stars: Mass-loss}

\begin{document}
\maketitle

\section{Binarity, Roche Model and basic definitions}

 The majority of the stars  are binaries or members of multiple systems.  Gravitational interaction in multiple systems makes possible detection of invisible bodies (from black holes to planets), test fundamental principles of physics (e.g. the case of the binary pulsar PSR 1913+16) and modify evolution of the components in the case of tight systems so much, that is necessary to define new classes of objects, like millisecond pulsars, microquasars and Cataclysmic Variables. Binary stars  are unique laboratories for testing stellar  models and theories of stellar evolution. The foundations of  modern astronomy is based on studies of gravitational bounded systems.

Since its origin, the Roche model (Roche 1973) has been a physical framework  to compare and test   binary star observations.  It assumes mass point stars, and originally considered only circular orbits. Any gas element in the co-rotating system "feels" an effective gravitational potential  known as the Roche potential, which includes the gravitational terms of the stellar components plus a centrifugal term. The relation between stellar sizes and Roche equipotential surfaces allow the classical definition of binaries in detached, semidetached and contact systems. The case of semi-detached binaries is especially interesting, since in that case a mass flow occurs from the secondary (the donor) to the primary (the gainer). During their evolution, a binary system can eventually pass for these three different configurations, although this is not always the case.
 
For evolution studies, conservative and non-conservative cases must be distinguished.
For the conservative case, the binary history depends basically on the ratio between their initial orbital period and the orbital period needed for contact, and the mass ratio (Eggleton 2006).  The contact could occur while the donor is having core H-burning (Case A), after  H exhaustion (Case B) or after  He exhaustion (Case C). In spite of being recognized as relevant for explaining the evolution of many binaries, non-conservative processes were rarely included in former evolution codes, 
but now are being considered as key elements in this kind of investigation 
(van Rensbergen et al. 2008).

\section{Interacting binaries, interaction modes and flow patterns}

\begin{small}
\begin{table*}
\centering
 \caption{Some types of interacting binaries.}
 \begin{tabular}{@{}ccccc@{}}  
Type& Usual configuration &Prototype &Typical orbit size  \\
& &(or example) & (in AU) &\\
\hline
Cataclysmic Variables & WD + RD  &U Gem&0.005\\
AM CVn & WD + WD &AM CVn&0.0003\\
Symbiotics &NS + giant/Mira + nebula &Z And &2\\
HMXRB &Be/O/B + BH/NS/WD &Cyg X-1 &0.05-2   \\
LMXRB &BH/NS + $<$ 1 M$\odot$  star&Her X-1& 0.001-0.2  \\
WR binaries & WR + O + nebula &$\gamma^{2}$ Vel     &0.1-12\\
Chromospherically active &fast rotator cool star  + companion &RS CVn  &0.1\\
W\,Serpentids, DPVs, Algols &MS + giant & $\beta$ Lyr, V\,393 Sco, $\beta$ Per&0.2\\
O or B  + O &MS/giants components &LSS\,3074   &0.1 \\
 \hline
\end{tabular}
\end{table*}
\end{small}

There are many types of interacting binaries (IBs), they are summarized in Table 1.  In the region of high density primaries  we have the High Mass X-Ray Binaries (HMXRB; Be/O/B + BH/NS/WD) along with the  microquasars,  these later characterized by spectroscopic evidence of relativistic jets. We have also the Low Mass X-Ray binaries
(LMXRB; BH/NS + lower mass companion, usually with mass  $<$  1M$\odot$) and the 
Cataclysmic Variable Stars (CVs), consisting of a white dwarf plus  a red dwarf. 
Magnetic braking and gravitational radiation drives CV evolution through the loss of  system angular momentum.  An open question in this field is the efficiency,  during the common envelope phase,
of energy removal from the system (Politano \& Weiler 2007, Davis et~ al. 2008).
The AM CVn stars consisting of 2 interacting WDs are other interesting class located in the very short orbital period tail. 
In the spectrum of lower density primaries we observe the chromospherically active binaries RS CVn,
 fast rotators with deep convective envelopes.  We also observe the Wolf Rayet binaries (WR + O + nebula) and the O + O pairs, sometimes interacting through their winds. 
Here we also have the W Serpentids, DPVs and Algols (MS +  giant, see \S~5).
On the long period side, we observe the Symbiotics, consisting of a NS plus a giant/Mira, they are usually surrounded by a nebula, probably feed by the red giant stellar wind.

Studies have shown that binaries interact through several mechanisms: (1)
Roche lobe overflow, implying gas streams, accretion discs, hot spots, gaseous environments, spiral shocks, (2) tidal friction between components, allowing orbit evolution/circularization and stellar deformation, or between one star and the disc, (3) gravitational radiation, especially when compact objects are present, (4) magnetic activity driven by rapid stellar rotation, (5) interaction of stellar winds in luminous stars, (6) reflection and irradiation phenomena, (7)
magnetic braking, especially in cool stars with convective inner region, (8) third body influence,
(9) fast spinning up of the primary through accretion from the donor and (10) mass and angular momentum loss.

Some flow patterns and physical features have been indirectly observed in IBs. The gas stream through the L1 point is revealed in line asymmetries, doppler maps and eclipse maps in semidetached binaries like CVs and Algols. The circumprimary accretion disc and interaction region is revealed in double peak emission, light curve humps, doppler maps and eclipse maps in CVs (Marsh \& Horne 1988, Marsh 2005). Circumprimary halo gas in WZ Sge  shows-up in IR photometry (Howell et~ al. 2008), and also has been predicted by simulations of close binary stars (Sytov et~ al.   2007).
The circumstellar matter in Algols is revealed by UV lines produced by photon scattering in induced polar winds (Plavec 1989). 
Doppler maps reveal magnetic channeled streams  in magnetic CVs (DQ Her systems and intermediate polars, Schwope 2001). Bipolar Jets appears like stationary non-eclipsed emission components in $\beta$ Lyrae and V\,356 Sgr (Ak et~ al. 2007, Polidan \& Lynch 1996) and dust shells have been detected in other W\,Serpentids (Taranova 1997). Outflows through L3 and L2 Lagrangian points have been detected in spectroscopy of RY Scuti (Djura\u{s}evi\'{c}, Vince \& Atanackovi\'{c} 2008) and OGLE\,05155332-6925581 (Mennickent et~ al. 2008, hereafter M08). 
Orbital period changes have been measured in $\beta$ Lyr (+19 s/year for 210 years indicating $\dot{M}$ $\approx$ 2$\times$10$^{5}$ M$\odot$/yr, Klimek \& Krimer 1973) and also in others W\,Ser  stars. 

\section{observational constraints}

The phenomena of humps, superhumps, late superhumps, outburst and outburst rebrightness, all of them  derived primarily from photometric monitoring,  place strong observational constrains on CV models (e.g. Osaki 1996). One of the more striking observational constraints in CVs of the SU UMa type is the correlation observed between the period excess and the orbital period (Stolz \& Schoembs 1984). The period excess is defined as (Ps-Po)/Po, where Ps is the superhump period and Po the orbital one. Models of superhumps have invoked disc precession driven by tidal instabilities to give account of this correlation (Murray 2000). Other interesting constraints are the spiral waves detected by Doppler tomography and eclipse mapping in dwarf novae during outburst and quiescence (Steeghs, Harlaftis \& Horne 1997) and recently the  
mini high states and oscillations detected in WZ Sge type CVs in quiescence (Zharikov et~ al. 2008).
 Retter, Richards \& Wu  (2005) reported superhumps in Algol and Doppler tomography by Richards (2007) reveals transient accretion
discs, hot spots and complex structures in similar systems. 
Interestingly, we observe similar phenomena at very different spatial scales in different types of IBs. 
The  superhumps observed in some HMXRBs and LMHRBs (Haswell et 
al. 2001a, 2001b), circumbinary discs suspected  in CVs, W\,Ser stars, DPVs and the disc-based symbiotic star outbursts, seems to support this "unified" picture, although it is clear that in some cases the systems are so complex that resist a simple interpretation.

\section{Insights from numerical simulations}

Simulations of the gas dynamics in IBs have helped to bring light 
on the processes occurring as result of interaction in a binary system. 
Early attempts clearly showed that gravity and tidal, centrifugal and viscous forces governed the  formation and evolution of the accretion disc, gas stream and interaction region (e.g. Lubow \& Shu 1975, Lubow 1989). 
Nowadays,  parallel supercomputers increase the chances of getting more realistic 
3D views of the systems. Sytov et~ al. (2008) modeled the gas dynamics in a typical CV system and found  several structures that can be associated with observational features, like accretion disc,
bow shock and spiral density waves. They also found a mechanism that could explain the mass losses from the binary. While several elements remains virtually statics in the binary system of rest during one orbital cycle, the bow shock position changes, allowing the escape of material through the L3 point twice per orbital period. This promising result could explain why matter is detected escaping  in the equatorial plane in some semidetached systems. 

\begin{figure}[]
\includegraphics[angle=0,width=\columnwidth,scale=1.1]{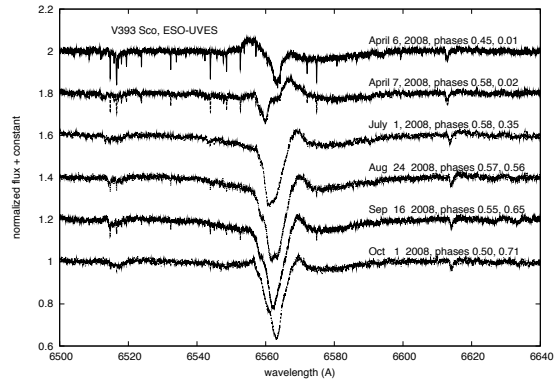}
\caption{Spectra of the Galactic DPV V\,393 Sco labeled with  observation  dates, orbital and long-cycle phases. H$\alpha$ emission is larger at long-cycle maximum. Orbital variability is illustrated in the two upper spectra. }
\end{figure}


\section{The W~Ser -- DPV -- Algol connection}

DPVs are intermediate mass, hot IBs showing two closely linked photometric periodicities (Mennickent et~ al. 2003, 2005, M08). A hundred of these systems have been found in the Magellanic Clouds and a dozen in our Galaxy (Mennickent \& Ko{\l}aczkowski 2009, hereafter M09). 
While the shorter period is the orbital one, the longer period has been interpreted as a cycle of mass depletion
into the interstellar medium (M08).  
DPVs show a constancy of the orbital period  in  time scales of a decade, contrary to $\beta$ Lyr. This constancy  in relatively high mass transfer rate systems has been explained as an equilibrium between mass exchange and mass loss in the long-term (M08). 
It is possible that high mass transfer rate spin-up the primary star until critical rotation is reached. This should  increase the disc mass and size until some instability enhances the mass loss into the circumbinary environment (at the maximum of the long periodicity). The larger emission line strength observed in the Galactic DPV V\,393 Sco near long-cycle maximum supports this conjecture (Fig.\,1).  
At this stage we can rise several interesting questions.
Have the period correlation observed in DPVs some physical connection with the correlation between superhump and orbital period found in CVs as proposed by M09? Is there an universal mass loss mechanism for semi-detached IBs at some specific evolutionary stages?  What is the contribution of these systems to the chemical enrichment of the ISM? Is the connection  W\,Serpentids
(massive/active) $\rightarrow$ DPVs (less massive/active) $\rightarrow$ Algols (even less massive/active) an evolutionary one or reveal different initial populations? This sequence probably illustrates how the mass transfer rate ($\dot{M_{2}}$) shapes the mass loss properties of the system (Table 2). In W\,Serpentids $\dot{M_{2}}$ could be so large that the mechanism triggering the DPV cycles does not work,  but matter escapes from the system chaotically, for instance through polar jets,  as observed in some W Serpentids.
Consequently in these systems we observe significant orbital period change ($\dot{P}$). In DPVs  $\dot{M_{2}}$ is just in the range needed to mantain a cycle of filling/depletion of the circumprimary disc. In DPVs  we observe  $\dot{P} \approx 0$ probably due to the balance between transferred and lost mass (M08).  Finally in Algols $\dot{M_{2}}$  is so small that produces small  mass loss and period changes. We note that the possible connection between W\,Serpentids and Algols was already noted by Wilson (1989); we have just added DPVs into a global picture. In our view the gainer rotates fast in W\,Ser stars and DPVs, but tidal forces
could have slowered rotation in the older Algols.



\begin{small}
\begin{table}
\centering
 \caption{Proposed qualitative description of W-Ser, DPVs and Algols.}
 \begin{tabular}{@{}cccc@{}}  
System& M and $\dot{M_{2}}$ &Mass loss, $\dot{P}$  &age  \\
\hline
W\,Ser &very large &large, large $\dot{P}$ &young \\
DPVs &large         &cyclic, $\dot{P} \approx 0$&middle\\
Algols&small         & small, $\dot{P} \approx 0$&old\\
 \hline
\end{tabular}
\end{table}
\end{small}

Since $\beta$ Lyr has been successfully observed interferometrically, revealing an asymmetrical gaseous  disc (Zhao et~ al. 2008) and
DPVs have roughly similar size, we propose that V\,393 Sco at $V$= 7.4 could be an additional target for interferometry. Dedicated interferometric studies could help us to understand  nonconservative evolution of IBs.

\begin{center}
DISCUSSION
\end{center}

\noindent 
{\it H. Sana}: A comment. In the list of types of IBs, you should add the O + O and O + B binaries that could also interact through the O star winds, similarly than WR + O binaries. \\ 
\noindent
\noindent
{\it A. Miroshnickenko:} Is there any circumstellar dust in Double Periodic Variables? \\ 
\noindent
{\it R.E. Mennickent:} We still do not know. We observe IR excess for many systems with 2MASS data,  and now are in the process of analyzing Spitzer data to answer this very important question.\\ 
\noindent
{\it W.J. de Wit:} Do the loops in the CMD go always clockwise? \\
 \noindent
{\it R.E. Mennickent:} We have only 1 system showing this behavior (M08), more analysis is needed to know if this is the rule for DPVs. \\
 \noindent
{\it P. Koubsk\'y:} A comment. Phase dependence of the H$\alpha$ emission does not mean interaction RLOF ($\phi$\,Per, HR 2142 show no evidence for RLOF). \\
 \noindent


\begin{thebibliography}

\bibitem[\protect\citeauthoryear{Ak et 
al.}{2007}]{2007A&A...463..233A} Ak, H., et al.\  2007, A\&A, 463, 233 


\bibitem[\protect\citeauthoryear{Davis et al.}{2008}]{2008MNRAS.389.1563D} 
Davis, P.J., Kolb, U., Willems, B., G{\"a}nsicke, B.T.\  2008, MNRAS, 389, 
1563 





\bibitem[\protect\citeauthoryear{Djura{\v s}evi{\'c}, Vince, 
\& Atanackovi{\'c}}{2008}]{2008AJ....136..767D} Djura{\v s}evi{\'c}, G., Vince, I., Atanackovi{\'c}, O.\, 2008, AJ, 136, 767 


\bibitem[\protect\citeauthoryear{Eggleton}{2006}]{2006epbm.book.....E} 
Eggleton, P.\, 2006, in Evolutionary Processes in Binary and Multiple Stars, Cambridge Astrophysics Series 40

\bibitem[\protect\citeauthoryear{Haswell et 
al.}{2001}]{2001MNRAS.321..475H} Haswell, C.~A., King, A.~R., Murray, J.~R., 
Charles, P.~A.\ 2001a, MNRAS, 321, 475 


\bibitem[\protect\citeauthoryear{Haswell et 
al.}{2001}]{2001ApSSS.276...41H} Haswell, C.~A., Rolfe, D.~J., King, A.~R., 
Murray, J.~R., Charles, P.~A.\  2001b, ApSSS, 276, 41 

\bibitem[\protect\citeauthoryear{Howell et al.}{2008}]{2008ApJ...685..418H} 
Howell, S.~B., Hoard, D.~W., Brinkworth, C., Kafka, S., Walentosky, M.~J., 
Walter, F.~M., Rector, T.~A.\, 2008, ApJ, 685, 418 


\bibitem[\protect\citeauthoryear{Klimek 
\& Kreiner}{1973}]{1973AcA....23..331K} Klimek, Z., Kreiner, J.~M.\ 1973, AcA, 23, 331 

\bibitem[\protect\citeauthoryear{Lubow}{1989}]{1989ApJ...340.1064L} Lubow, 
S.~H.\ 1989, ApJ, 340, 1064 

\bibitem[\protect\citeauthoryear{Lubow 
\& Shu}{1975}]{1975ApJ...198..383L} Lubow, S.~H., Shu, F.~H.\ 1975, ApJ, 198, 383 


\bibitem[\protect\citeauthoryear{Marsh}{2005}]{2005Ap&SS.296..403M} Marsh, T.~R.\ 2005, Ap\&SS, 296, 403 

\bibitem[\protect\citeauthoryear{Marsh 
\& Horne}{1988}]{1988MNRAS.235..269M} Marsh, T.~R., Horne, K.\ 1988, MNRAS, 235, 269 


\bibitem[\protect\citeauthoryear{Mennickent et 
al.}{2003}]{2003A&A...399L..47M} Mennickent, R.~E., Pietrzy{\'n}ski, G., Diaz, M., Gieren, W.\ 2003, A\&A, 399, L47 

\bibitem[\protect\citeauthoryear{Mennickent et 
al.}{2005}]{2005MNRAS.357.1219M} Mennickent R.~E., Cidale L., D{\'{\i}}az 
M., Pietrzy{\'n}ski G., Gieren W., Sabogal B., 2005, MNRAS, 357, 1219 

\bibitem[\protect\citeauthoryear{Mennickent et 
al.}{2008}]{2008MNRAS.389.1605M} Mennickent, R.~E., Ko{\l}aczkowski, Z., 
Michalska, G., Pietrzy{\'n}ski, G., Gallardo, R., Cidale, L., Granada A., 
Gieren, W.\ 2008, MNRAS, 389, 1605 

\bibitem[\protect\citeauthoryear{Mennickent and Kolaczkowski}{2009}]{2009} Mennickent, R.E. \& Ko{\l}aczkowski, Z.\  2009, in
proceedings of the XII Reuni\'on Latinoamericana de la IAU, \RMAA, in press (arXiv:0903.4819)


\bibitem[\protect\citeauthoryear{Murray}{2000}]{2000MNRAS.314L...1M} Murray, 
J.~R.\ 2000, MNRAS, 314, L1 

\bibitem[\protect\citeauthoryear{Osaki}{1996}]{1996PASP..108...39O} Osaki, 
Y.\ 1996, PASP, 108, 39 

\bibitem[\protect\citeauthoryear{Plavec}{1989}]{1989SSRv...50...95P} Plavec, 
M.~J.\ 1989, SSRv, 50, 95 

\bibitem[\protect\citeauthoryear{Polidan 
\& Lynch}{1996}]{1996AAS...188.3804P} Polidan, R.~S., Lynch, D.~E.\ 1996, AAS, 28, 881 

\bibitem[\protect\citeauthoryear{Politano 
\& Weiler}{2007}]{2007ApJ...665..663P} Politano, M., Weiler, K.~P.\ 2007, ApJ, 665, 663 

\bibitem[\protect\citeauthoryear{Retter, Richards, 
\& Wu}{2005}]{2005ApJ...621..417R} Retter, A., Richards, M.~T., Wu, K.\, 2005, ApJ, 621, 417 

\bibitem[\protect\citeauthoryear{Richards}{2007}]{2007IAUS..240..160R} 
Richards, M.~T.\ 2007, Proceedings of the IAU, 240, 160 

\bibitem[\protect\citeauthoryear{Roche}{1873}]{} Roche, E.\ 1873, Mem. Acad. Sci. Montpellier 8, 235 


\bibitem[\protect\citeauthoryear{Schwope}{2001}]{2001LNP...573..127S} 
Schwope, A.\  2001, Lecture Notes in Physics, 573, 127 

\bibitem[\protect\citeauthoryear{Stolz \& Schoembs}{1984}]{1984A&A...132..187S} Stolz, B., Schoembs, R.\ 1984, A\&A, 132, 187 

\bibitem[\protect\citeauthoryear{Steeghs, Harlaftis, 
\& Horne}{1997}]{1997MNRAS.290L..28S} Steeghs, D., Harlaftis, E.~T., Horne, K.\ 1997, MNRAS, 290, L28 


\bibitem[\protect\citeauthoryear{Sytov et al.}{2007}]{2007ARep...51..836S} 
Sytov, A.~Y., Kaigorodov, P.~V., Bisikalo, D.~V., Kuznetsov, O.~A., Boyarchuk, 
A.~A.\, 2007, ARep, 51, 836 

\bibitem[\protect\citeauthoryear{Taranova}{1997}]{1997AstL...23..704T} 
Taranova, O.~G.\ 1997, AstL, 23, 704 



\bibitem[\protect\citeauthoryear{van Rensbergen et 
al.}{2008}]{2008A&A...487.1129V} van Rensbergen, W., de Greve, J.~P., de Loore C., Mennekens, N.\ 2008, A\&A, 487, 1129 




\bibitem[\protect\citeauthoryear{Wilson}{1989}]{1989SSRv...50..191W} Wilson, 
R.~E.\ 1989, SSRv, 50, 191 

\bibitem[\protect\citeauthoryear{Zhao et al.}{2008}]{2008ApJ...684L..95Z} 
Zhao, M., et al.\, 2008, ApJ, 684, L95 

\bibitem[\protect\citeauthoryear{Zharikov et 
al.}{2008}]{2008A&A...486..505Z} Zharikov, S.~V., et al.\  2008, A\&A, 486, 505 



\end{thebibliography}
\end{document}